# Substrate free synthesis of graphene nanoflakes by atmospheric pressure chemical vapour deposition using Ni powder as a catalyst

JOYDIP SENGUPTA[1,*], KAUSTUV DAS[2], U N NANDI[3] and CHACKO JACOB[4]

[1]Department of Electronic Science, Jogesh Chandra Chaudhuri College, Kolkata 700033, India
[2]Department of Physics, Jadavpur University, Kolkata 700032, India
[3]Department of Physics, Scottish Church College, Kolkata 700006, India
[4]Materials Science Centre, Indian Institute of Technology, Kharagpur 721302, India
*Author for correspondence (joydipdhruba@gmail.com)

**Abstract.** Graphene nanoflakes (GNFs) were synthesized by atmospheric pressure chemical vapour deposition of propane ($C_3H_8$) employing Ni (salen) powder without the introduction of a substrate. The graphitic nature of the GNFs was examined by an X-ray diffraction method. Scanning electron microscopy results revealed that GNFs were stacked on top of one another and had a high aspect ratio. Transmission electron microscopy studies suggested that the GNFs were made up of a number of crystalline graphene layers, some of which were even single crystalline as evident from the selected area diffraction pattern. Finally, Raman spectroscopy confirmed the high quality of the GNFs.

## 1. Introduction

Graphene is an atomically thin carbon nanomaterial made up of a single sheet of $sp^2$ hybridized carbon atoms. It has a unique nanostructure with fascinating chemical and physical properties [1] and is ideally suited for further miniaturizing electronics to form ultra-small-scale devices. It is used in numerous applications such as nanoelectronics [2], energy storage [3] and sensors [4]. To meet these essential requirements in technological fields, the synthesis of low cost and large-area good quality graphene is highly desirable. Apart from the first successful fabrication of graphene by mechanical exfoliation of graphite [5], the other reported processes for the synthesis of graphene include chemical exfoliation of graphite [6], thermal decomposition of SiC [7], chemical vapour deposition (CVD) of hydrocarbons [8], epitaxial growth on the substrate surface [9] and solution-based chemical reduction of graphene oxide into graphene [10]. Though a number of routes exist for the synthesis of graphene, most of them are not well-suited for integration with current silicon technology. Among the different synthesis routes, CVD has gained popularity for the production of graphene due to its economical and scalable approach along with its compatibility with current silicon technology. Many researchers have employed atmospheric/ambient pressure CVD to synthesize graphene using either Cu [11–13] or Ni [14–17] as a catalyst. The Ni-catalysed atmospheric pressure chemical vapour deposition (APCVD) synthesis of graphene generally employed either single or polycrystalline Ni thin films deposited on a suitable substrate. In general, the process of Ni film deposition suffers from two major drawbacks. Firstly, it requires ample amounts of time and secondly, a uniform deposition area is indeed limited. These drawbacks can be surmounted by using the Ni catalyst directly in a powder form within the reactor chamber for the growth of graphene. Moreover, to integrate graphene into device prototypes, the transfer of graphene from the substrate is a mandatory step which adds complexity to the entire process [18]. Such limitations of the existing graphene synthesis methods point to the necessity of a process which will be free from the use of the substrate, thereby removing the 'transfer' step and reducing the production cost. Subsequently, it was also suggested that the $H_2$-treated Ni surface is beneficial to obtain relatively defect-free graphene *via* the CVD process [19]. Other than the APCVD process, researchers have also used microwave-assisted CVD [20], plasma-enhanced CVD [21], microwave plasma-enhanced CVD [22,23] and even a chemical route [24] for the synthesis of graphene which include both catalytic and non-catalytic substrates for graphene growth.

In this paper, an efficient and facile process to synthesize graphene nanoflakes (GNFs) using APCVD of propane employing Ni powder without the introduction of the substrate is described. Scanning electron microscopy (SEM) and transmission electron microscopy (TEM) were employed to characterize the overall growth morphology and the internal structure of the GNFs, respectively. Information regarding the crystallographic structure and chemical composition was obtained from X-ray diffraction (XRD) and Raman spectroscopy.

## 2. Experimental

Ni (salen) powder (yellow in colour) was poured into a quartz boat and inserted into the quartz tube furnace as shown in figure 1. At first, the chamber was purged with Ar up to atmospheric pressure and then the Ni (salen) powder was heated with a heating rate of 6°C min$^{-1}$ (approx.) in a Ar atmosphere with a flow rate of 1 standard litre per minute (slm) until 900°C. Thereafter, Ar was replaced with $H_2$ with a flow rate of 1 slm and the Ni (salen) powder was further annealed in $H_2$ for 10 min. Finally, the furnace temperature was reduced to 850°C and $H_2$ was replaced with $C_3H_8$ with a flow rate of 200 sccm for 1 h to initiate the growth process. On completion of the growth period, $C_3H_8$ was turned off and $H_2$ with a flow rate of 1 slm, was introduced into the reactor. The reactor was allowed to cool down with a cooling rate of 3°C min$^{-1}$ (approx.) to room temperature in a $H_2$ atmosphere to avoid any oxidation of the samples. Finally, the flow of $H_2$ was cut off and the chamber was purged with Ar before it was opened. After the completion of the growth procedure, a black-coloured powder was collected from the quartz boat and used for further characterization. XRD (Bruker D8 powder) with a Cu source was used to characterize the crystalline nature of GNFs. SEM (FEI Inspect F-50 and MIRA3 TESCAN) and TEM (Tecnai G2 20 S-TWIN) were used to obtain information regarding the overall growth morphology and the internal microstructure of the GNFs. A 488 nm air-cooled Ar$^+$ laser was used as an excitation source for Raman measurements at room temperature to test the quality of the GNFs.

## 3. Results and discussion

Figure 2 shows the XRD pattern of GNFs at room temperature. It is clearly observed from the figure that the GNFs exhibited a sharp peak with very high intensity centred at 26.45°. This peak in the XRD spectrum corresponds to the (002) plane of the hexagonal graphite structure and no turbostratic graphite peak was noticed [25]. The position of the peak was used to calculate the inter-planar distance between two adjacent graphene sheets and the extracted value was found to be 3.36 Å which matches well with the inter-planar spacing corresponding to the (002) plane of the hexagonal graphite structure. From the observed peak broadening with the full width at half maximum of 0.42, it could be inferred that the stacking of graphene was well ordered.

SEM was used to analyse the morphology of the GNFs. The SEM micrographs (figure 3a and b) showed an ensemble of flaky structures with a high aspect ratio (width-to-thickness) and stacked on top of each other. Considerable amount of curved and corrugated edges were observed in these GNFs. SEM analysis indicated that the grown material consisted of multi-layered graphene sheets with a flaky morphology.

The internal structure of GNFs was examined using TEM measurement. The sheet-like nature of the flakes was clearly apparent from the TEM micrographs (figure 4a and b) revealing that the graphene sheets to be few layers thick. The TEM image (figure 4a) from the edge of the flakes indicated that the number of layers to be five or less. It could also be noticed that the edges of the GNFs were folded back in some areas (figure 4b). Moreover, these GNFs were highly transparent and extremely stable under the exposure of the electron beam.

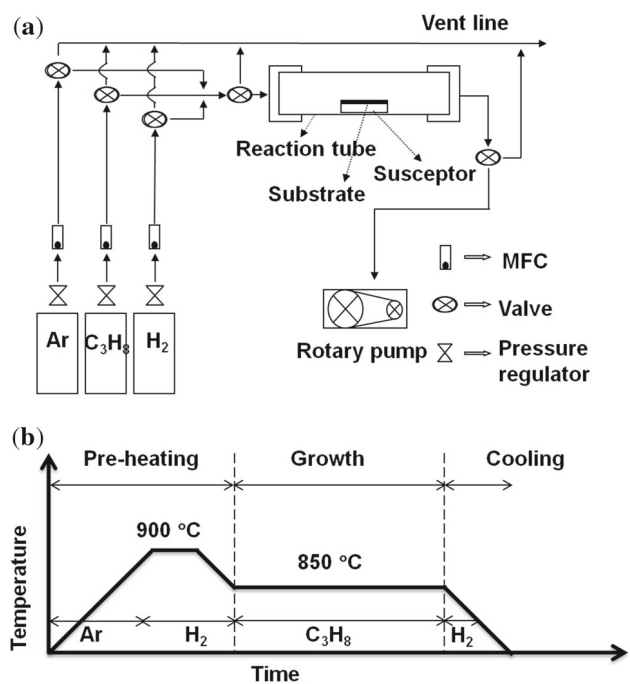

**Figure 1.** (a) Schematic representation of the APCVD system and (b) temperature–time graph of graphene synthesis *via* APCVD (not to scale).

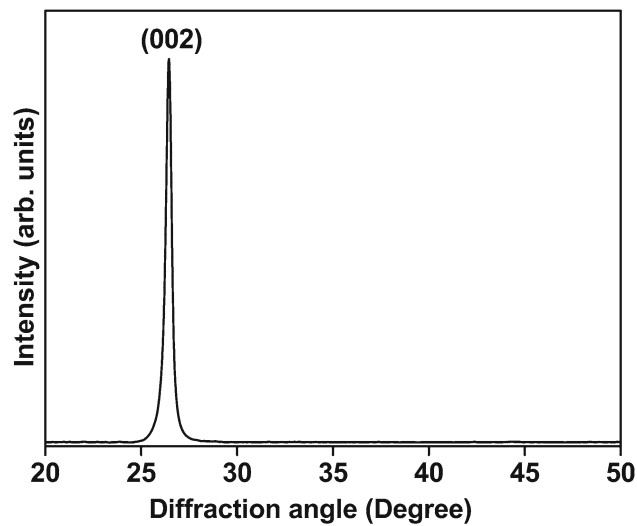

**Figure 2.** XRD spectrum of GNFs at room temperature.

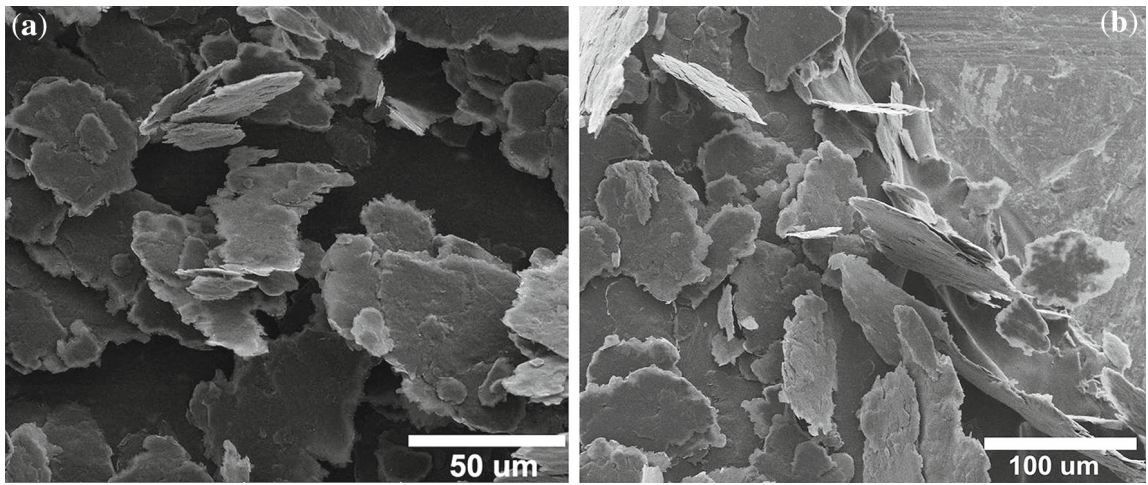

**Figure 3.** (**a** and **b**) SEM micrographs of the as-grown GNFs deposited using APCVD.

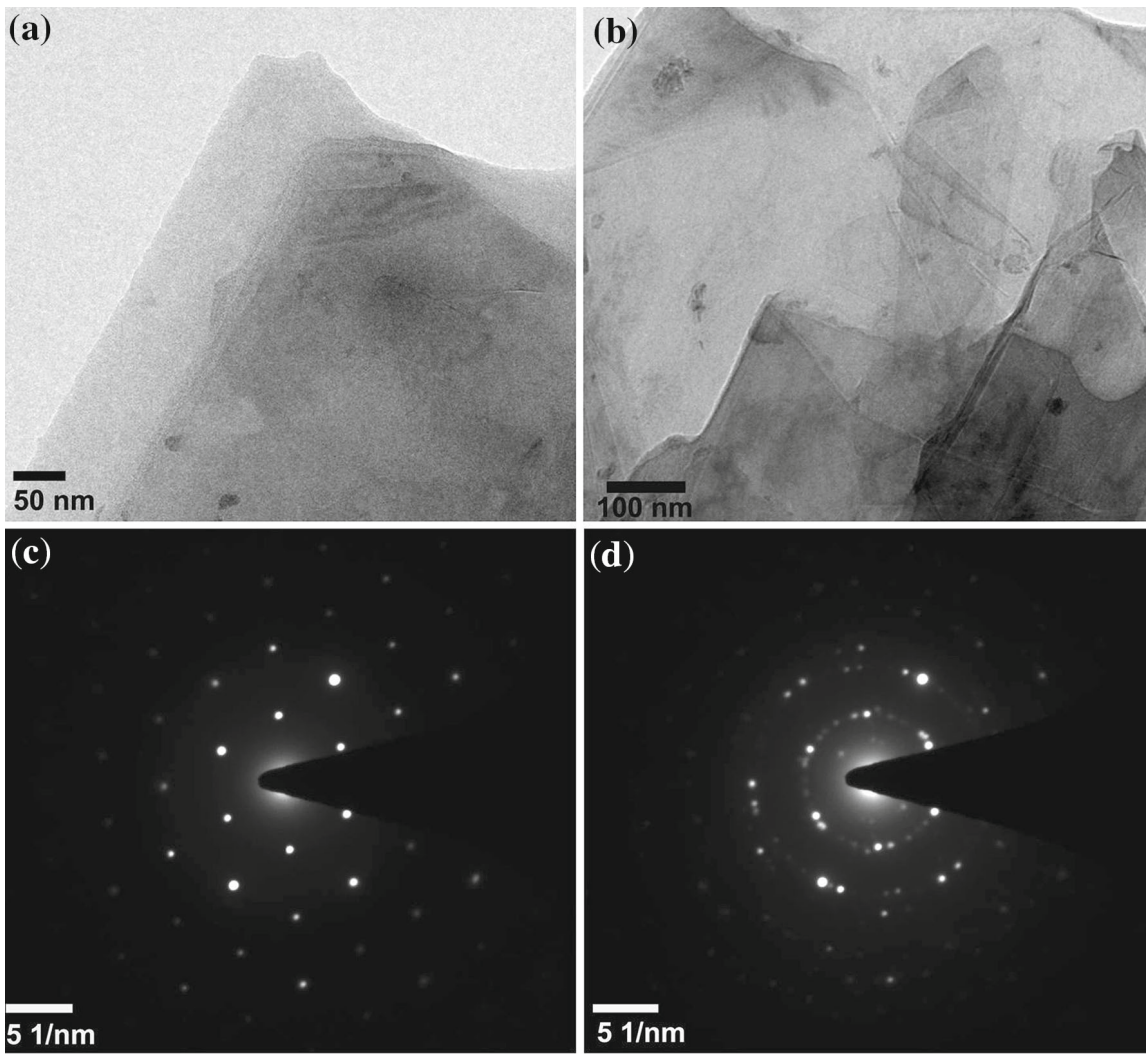

**Figure 4.** (**a** and **b**) TEM images of the GNFs grown by the APCVD method using Ni, SAED pattern obtained (**c**) from the edge of the flake and (**d**) from a region of few hundreds of nm inside the flake.

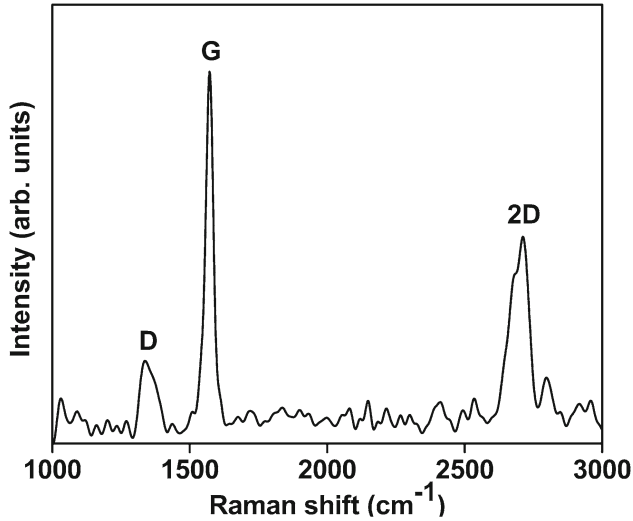

**Figure 5.** Raman spectrum (488 nm excitation) of GNFs at room temperature.

The crystalline nature of the GNFs was also evidenced by the normal-incidence selected area electron diffraction (SAED) patterns shown in figure 4c and d. The SAED pattern obtained from the edge of a flake (figure 4c) displayed a well-defined hexagonal diffraction spot pattern which is the signature of single layer graphene. Such spots instead of rings implied that the flakes did not contain randomly oriented domains. However, the SAED pattern (figure 4d) obtained from a region of few hundreds of nm inside the flake revealed concentric incomplete rings instead of individual spots which implied that the individual graphene layers were randomly stacked on top of one another. Each individual sheet contributed to the diffraction pattern causing the spots to merge into incomplete rings.

Raman spectrum (figure 5) contained three major bands denoted as D, G and 2D. The D band at 1337 cm$^{-1}$ arose from the defect-mediated zone-edge (near K-point) phonons. These defects may be caused by the wrinkled, non-uniform, twisted and folded regions of GNFs [26]. SEM analysis revealed that the grown GNFs had a substantial amount of twisted and wrinkled regions, while folded regions of GNFs were noticed in TEM micrographs. These microscopic findings are in agreement with the evolution of the D peak. The peak at 1572 cm$^{-1}$ represents the G band originating due to the doubly degenerate $E_{2g}$ mode, and shows the presence of crystalline graphitic carbon in GNFs. This result was consistent with XRD analysis. The second order of the D band, the so-called 2D band, at around 2711 cm$^{-1}$ evolved from second-order double resonant Raman scattering.

The intensity ratio of the D peak to the G peak was 0.22, suggesting that the grown GNFs were crystalline. To evaluate the in-plane sp$^2$ crystallite size ($L_a$) for the nanographitic material, the formula suggested by Cançado et al [27] was used where the laser line wavelength ($\lambda_l$) was in nanometre units

$$L_a \text{ (nm)} = \left(2.4 \times 10^{-10}\right) \lambda_l^4 \frac{I_G}{I_D}. \quad (1)$$

The value of $L_a$ calculated from the above formula was found to be 61 nm. Moreover, the intensity ratio of the 2D peak to the G peak was 0.55 which indicated that the grown material belongs to the few layer graphene (FLG) system [28]. Similar results were also obtained by Zhang et al [14] for the graphene sample grown by APCVD using polycrystalline Ni film. However, they deposited Ni metal by e-beam evaporation on a SiO$_2$/Si wafer which is much more complex than the direct use of Ni powder.

Because of the finite solubility of carbon in Ni [29], Ni is widely used as a catalyst [30] for the synthesis of graphene in CVD. In the CVD method, a carbon source is decomposed at high temperature when it is in contact with the Ni surface. Thereafter, the decomposed carbon atoms initially diffuse into Ni, but finally segregate and precipitate at the surface of Ni to form graphene [31]. It was further reported that the atomically smooth substrate surface and the absence of grain boundaries (e.g., single crystalline Ni) are favourable conditions for the formation of monolayer/bilayer graphene. Whereas, the presence of grain boundaries in the substrate (e.g., polycrystalline Ni) serve as nucleation sites for multilayer graphene growth [14]. Here, Ni powder was used for graphene growth and at the growth temperature, it may act as polycrystalline Ni which leads to the growth of few layered graphene. However, rigorous experiments need to be carried out to confirm the exact growth mechanism of graphene using Ni powder.

## 4. Conclusion

In this paper, a facile and efficient method for the simple and scalable synthesis of GNFs using Ni(salen) powder without the introduction of a substrate has been described with highly reproducible features. GNFs were synthesized by the APCVD method and characterized using standard techniques such as XRD, SEM, TEM and Raman. The graphitic nature of GNFs was predicted from the XRD result. SEM and TEM analyses confirmed that the grown material consists of multilayered graphene sheets with a flaky morphology. Raman spectroscopy confirmed the good quality of the sample and suggested that the sample belonged to the FLG system. This simple, scalable, efficient and low cost (since no substrate is required) technique for the synthesis of GNFs will open up scope for further exploration in the subject.

**Acknowledgements**

This work was supported by the Science and Engineering Research Board, Department of Science & Technology, India, under project number (SR/FTP/PS-120/2012) and (SB/S2/CMP-099/2013). We are grateful to Dr A Singha from the Department of Physics, Bose Institute, for his help with Raman spectroscopy. We would also like to thank Dr P Majhi from the Department of Polymer and Process Engineering, IIT Roorkee, for his help with SEM characterization.

**References**

[1] Geim A K and Novoselov K S 2007 *Nat. Mater.* **6** 183
[2] Westervelt R M 2008 *Science* **320** 324
[3] Pumera M 2011 *Energy Environ. Sci.* **4** 668
[4] Justino C I L, Gomes A R, Freitas A C *et al* 2017 *Trends Anal. Chem.* **91** 53
[5] Novoselov K S, Geim A K, Morozov S V *et al* 2004 *Science* **306** 666
[6] Ciesielski A and Samorì P 2014 *Chem. Soc. Rev.* **43** 381
[7] Mishra N, Boeckl J, Motta N *et al* 2016 *Phys. Status Solidi (a)* **213** 2277
[8] Wang X, You H, Liu F *et al* 2009 *Chem. Vapor Depos.* **15** 53
[9] Tetlow H, Boer J P, Ford I J *et al* 2014 *Phys. Rep.* **542** 195
[10] Chen W, Yan L and Bangal P R 2010 *J. Phys. Chem. C* **114** 19885
[11] Qaisi R M, Smith C E and Hussain M M 2014 *Phys. Status Solidi RRL* **8** 621
[12] Wang S, Hibino H, Suzuki S *et al* 2016 *Chem. Mater.* **28** 4893
[13] Nguyen V T, Le H D, Nguyen V C *et al* 2013 *Adv. Nat. Sci: Nanosci. Nanotechnol.* **4** 035012
[14] Zhang Y, Gomez L, Ishikawa F N *et al* 2010 *J. Phys. Chem. Lett.* **1** 3101
[15] Fogarassya Z, Rümmeli M H, Gorantla S *et al* 2014 *Appl. Surf. Sci.* **314** 490
[16] Reina A, Jia X, Ho J *et al* 2009 *Nano Lett.* **9** 30
[17] Reina A, Thiele S, Jia X *et al* 2009 *Nano Res.* **2** 509
[18] Avouris P and Dimitrakopoulos C 2012 *Mater. Today* **15** 86
[19] Losurdo M, Giangregorio M M, Capezzuto P *et al* 2011 *Phys. Chem. Chem. Phys.* **13** 20836
[20] Sun Y, Yang L, Xia K *et al* 2018 *Adv. Mater.* **30** 1803189
[21] Malesevic A, Vitchev R, Schouteden K *et al* 2008 *Nanotechnology* **19** 305604
[22] Bo Z, Yang Y, Chen J *et al* 2013 *Nanoscale* **5** 5180
[23] Shang N G, Papakonstantinou P, McMullan M *et al* 2008 *Adv. Funct. Mater.* **18** 3506
[24] Yoon S M, Choi W M, Baik H *et al* 2012 *ACS Nano* **6** 6803
[25] Soin N, Roy S S, O'Kane C *et al* 2011 *CrystEngComm* **13** 312
[26] Meyer J C, Kisielowski C, Erni R *et al* 2008 *Nano Lett.* **8** 3582
[27] Cañado L G, Takai K, Enoki T *et al* 2006 *Appl. Phys. Lett.* **88** 163106
[28] Marchena M, Song Z, Senaratne W *et al* 2017 *2D Mater.* **4** 025088
[29] Natesan K and Kassner T F 1973 *Metall. Trans.* **4** 2557
[30] Seah C M, Chai S P and Mohamed A R 2014 *Carbon* **70** 1
[31] Li X, Cai W, Colombo L *et al* 2009 *Nano Lett.* **9** 4268